\documentclass[11pt,a4paper]{article}

\usepackage[latin1]{inputenc}
\usepackage{amsmath}
\usepackage{amsfonts,amssymb,graphicx,amsthm,listings,float}
\usepackage{caption,subcaption,color,chngcntr,cite,authblk,bbold,braket}
\usepackage[titletoc]{appendix}  
\usepackage[english]{babel} 
\usepackage[left=2.00cm, right=2.00cm, top=2.00cm, bottom=2.00cm]{geometry}

\usepackage{soul}
\usepackage{xcolor}

\title{Design of quasiperiodic magnetic superlattices and domain walls supporting bound states}

\author{Miguel Castillo-Celeita\footnote{mfcastillo@fis.cinvestav.mx} }
\author{Alonso Contreras-Astorga\footnote{alonso.contreras@conacyt.mx}}
\author{David J. Fern\'andez C.\footnote{david.fernandez@cinvestav.mx}}

\affil{\small $~^{* \ddag}$ Physics Department, Cinvestav, P.O. Box. 14-740, 07000 Mexico City, Mexico \\ 
$~^\dag$ CONAHCyT-Physics Department, Cinvestav, P.O. Box. 14-740, 07000 Mexico City, Mexico}

\date{}
\begin{document}
\maketitle

\begin{abstract}
We study the simplest Lam\'e magnetic superlattice in graphene, finding its allowed and forbidden energy bands and band-edge states explicitly. Then, we design quasiperiodic magnetic superlattices supporting bound states using Darboux transformations. This technique enables us to add any finite number of bound states, which we exemplify with the most straightforward cases of one and two bound states in the designed spectrum. The topics of magnetic superlattices and domain walls in gapped graphene turn out to be connected by a unitary transformation in the limit of significantly large oscillation periods. We show that the generated quasiperiodic magnetic superlattices are also linked to domain walls, with the bound states keeping their nature in such a limit.        
\end{abstract}

\section{Introduction}

\emph{Graphene} is a two-dimensional material composed of a single layer of carbon atoms placed in a hexagonal arrangement with remarkable mechanical and electrical properties. At low energy, its charge carriers are described as massless Dirac electrons with constant Fermi velocity \cite{Wallace1947,CastroNeto2009}. For this reason, this material can be used to perform bench-top experiments of analogs of quantum electrodynamics \cite{Semenoff1984}. The quantum Hall effect \cite{Novoselov2005}, Klein tunneling \cite{Katsnelson2006,stander2009,Young2009},  Zitterbewegung \cite{Katsnelson2006b, Rusin2008}, and Schwinger mechanism \cite{Allor2008,Fillion-Gourdeau2015} are examples of relativistic phenomena studied in graphene. 

Let us notice that a superlattice appears when graphene is placed in periodic external fields or interactions. The study of periodic systems goes back to 1931 when Kronig and Penney introduced a model of electrons in crystal lattices in non-relativistic quantum mechanics \cite{Kronig1931}.
In the context of graphene, we can find nowadays in the literature electrostatic \cite{Park2008,Barbier2008,Park2009,Brey2009} or magnetic \cite{ibrapeet95,life06,wu2008,DellAnna11} superlattices; there are also gapped graphene superlattices where there is a mass term involved \cite{Maksimova2012}. Such fluctuations can be caused directly by periodic external electromagnetic fields or by the deposition of a monolayer graphene on specific substrates. When graphene is deposited on such substrates, the symmetry in the hexagonal lattice can be broken if electrons on a triangular sublattice  have different hopping energy  to the other sublattice. This behavior is observed for Boron-Nitride: this material has the same graphene lattice structure with two different types of atoms breaking down the symmetry \cite{Correa13}. Epitaxial graphene growth on a Silicon-Carbide substrate also allows obtaining the same behavior \cite{Semenoff08,gafi21,nataf20, ghosh2021,roy2021,schulze2022}. A mass term in the Dirac equation models such asymmetry and produces a spectral gap. An interesting limit occurs when the oscillation period of the mass term is large: if we get a mass-type barrier, it is said that we deal with domain walls \cite{Semenoff08}.   

On the other hand, supersymmetric quantum mechanics (SUSY) for physicists, or Darboux transformation (DT) for mathematicians, is an algorithm to expand the family of solvable potentials in non-relativistic quantum mechanics by using a known solvable Schr\"odinger equation as an input. In general, the initial Hamiltonian and its SUSY partners share some similarities, such as the asymptotic behavior of their potentials and their domain of definition. The corresponding spectra differ only in a finite number of energy levels. In particular, the Schr\"odinger equation for periodic potentials and its supersymmetric extensions, or SUSY partners, has been studied in \cite{Dunne1998,Khare1999,fnn00,Fernandez2002b,Fernandez2002,fg05,fg07,bermudez2012,bermudez2013}. Moreover, the SUSY technique has been successfully adapted to the Dirac equation in different settings, see for example  \cite{kuru2009,yesilta2012,midya2014,Contreras-Astorga2014,junker2020,celeita2020,bagchi2021,kizilirmak2021,schulze2021}. The introduction of magnetic or pseudo-magnetic fields modifies the spectrum and may confine the electrons into a finite space region \cite{raya2010,naba17,diaz2020,yesilta2022}. 

This work aims to address a Lam\'e magnetic superlattice in graphene, an exactly solvable system whose spectrum is composed of allowed and forbidden energy bands. Moreover, we will analyze how to insert bound states using Darboux transformations. The technique will introduce aperiodicities in the superlattice, which are responsible for the inserted bound states. Furthermore, by means of a unitary rotation, we will connect the problem of graphene in an external magnetic field with the problem of gapped graphene, which is modeled by adding a position-dependent mass term in the Dirac equation. 
An interesting limit is obtained for the period of the superlattice going to infinity, when the rotated system converts into a domain wall which keeps the added bound states.     

\section{Lam\'e magnetic superlattices}\label{sec Lame}
Low energy charge carriers in graphene satisfy the Dirac equation $H \Psi = v_F \Vec{\sigma} \cdot \mathbf{p} \Psi = E \Psi$, where $\mathbf{p}$ is the momentum operator, $\Vec{\sigma}=(\sigma_1, \sigma_2)$ is a vector array whose components are the Pauli matrices, and $v_F \approx c/300$ is called Fermi velocity, and we placed the graphene layer in the $x-y$ plane.  To introduce the effect of a magnetic field perpendicular to the graphene layer $\mathbf{B}= B_z \hat{\textbf{k}} = \nabla \times \mathbf{A}$ we use Peierls substitution (or minimal coupling) $\mathbf{p} \rightarrow \mathbf{p}+e \mathbf{A}$; we restrict ourselves to problems with translational invariance along $y-$axis and use the Landau gauge, $\mathbf{A}=(0,A_y(x),0)$; thus the wavefunction can be written as $\Psi(x,y)= e^{i k_y y} \psi(x)$. Then, the 2-dimension (2D) Dirac equation reduces to the 1-dimension (1D) one: 
\begin{equation}
\left(-i \sigma_1 \partial_x + k_y \sigma_2 + \frac{e}{\hbar} A_y \sigma_2 \right) \psi = \frac{E}{\hbar v_F} \psi.     
\end{equation}
To obtain a dimensionless equation, we introduce the magnetic length $\ell_B = \sqrt{\hbar  /e B_0}$, express $A_y= B_0 \ell_B (-\ell_B k_y+W_0(x))$ where $B_0$ is the magnetic field strength, and replace $x \rightarrow x/\ell_B, ~k_y \rightarrow \ell_B k_y,~ E \rightarrow \ell_B E/ \hbar v_F$, resulting in  
\begin{equation}
H \psi= \left[-i \sigma_1 \partial_x +  W_0 \sigma_2 \right] \psi =  E \psi.     \label{Dirac Mag}
\end{equation}
With the previous replacements the length $x$ and momentum $k_y$ are measured in units of $\ell_B$ and $\ell_B^{-1}$, respectively, and the energy in units of $\ell_B/\hbar v_F$. Typical values of magnetic field strengths $B_0$ in ferromagnetic materials range from $0.1$T to $1$T, which corresponds with $\ell_B\approx 80$nm - $25$nm \cite{DellAnna11}. As a remark,  if a Hamiltonian $H$ as in \eqref{Dirac Mag} has an eigenvector $\psi_{E^+}$ with eigenvalue $E^+\geq 0$ in the positive part of the spectrum, then $\psi_{E^-} = \sigma_3 \psi_{E^+}$ is an eigenvector with energy $E^- = - E^+$. 

Since $\psi(x)=(\psi^+(x), \psi^-(x))^T$ is a two-entry spinor, we can write \eqref{Dirac Mag} as a coupled system of equations: 
\begin{align}
 - i \psi^-_x - i W_0\psi^-    & = E \psi^+,  \nonumber  \\
 - i \psi^+_x + i W_0\psi^+    & = E \psi^-. \label{coupled}  
\end{align}
Throughout this paper, we will use the notation $f_x = \partial_x f(x)$. Such a system can be decoupled by inserting one of these equations into the other and vice versa, resulting in 
\begin{equation}
 h^\pm \psi^\pm = \varepsilon \psi^\pm, \label{dirac schrodinger}
\end{equation}
where $h^\pm =- \partial_x^2 + V^\pm(x)$, $V^\pm= W_0^2 \pm W_{0 x}$, and $\varepsilon=E^2$. 

\begin{figure}[t]
\centering
 \includegraphics[width=0.4\textwidth]{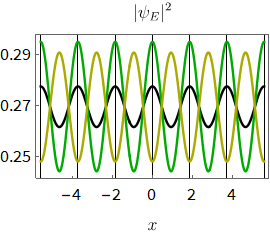}
 \caption{Lam\'e magnetic superlattices. Probability densities of the band edge-states,  $|\psi_{E_0}|^2$ (black line), $|\psi_{E_{\sqrt{1-m}}}|^2$ (green line), $|\psi_{E_1}|^2$ (yellow line) for $m=1/2$ and $A_1=A_2=1$.}\label{Lame states}
\end{figure}

Now, let us introduce the simplest Lam\'e magnetic superlattice and the  corresponding vector potential amplitude
\begin{equation} \label{lame magnetic}
    B_z(x)=B_0 \frac{m-1+ \text{dn}(x|m)^4}{\text{dn}(x|m)^2}  ,\qquad W_0(x)=m \frac{\text{cn}(x|m) \text{sn}(x|m)}{\text{dn}(x|m)},
\end{equation}
where $m$ is the modulus of the Jacobi elliptic functions $\text{sn}(x|m),~ \text{cn}(x|m),~\text{dn}(x|m)$, such that $0 \leq m \leq 1$. The two Schr\"odinger potentials  $V^- = -m+2m ~{\text{sn}^2(x|m)}$ and $V^+ = -m+2m ~{\text{sn}^2(x+K|m)}$ are called Lam\'e potentials, whose period is $\mathcal{T}=2K(m)$, where $K(m)=\int_0^{\pi/2} (1-m \sin^2\theta)^{-1/2}d\theta$.
By solving the Schr\"odinger equation $h^- \psi^- = \varepsilon \psi^-$ (see Appendix \ref{Ap Lame} for more details) and using \eqref{coupled} we can obtain the eigenspinors and the allowed and forbidden bands of the Dirac problem.  The allowed positive energies are then $E^+ = [0, \sqrt{1-m}] \cup [1, \infty)$. Note that the spectrum does not depend on the wavenumber $k_y$. The positive band-edge states are 
\begin{align}
 \psi_{E_0}  = \begin{pmatrix} -i A_1 ~\text{dn}(x+K|m)  \\ A_2~ \text{dn}(x|m) \end{pmatrix}, \quad 
  \psi_{E_{\sqrt{1-m}}}  = \begin{pmatrix} -i ~\text{cn}(x+K|m) \\ \text{cn}(x|m) \end{pmatrix}, \quad 
  \psi_{E_1} =\begin{pmatrix} -i~ \text{sn}(x+K|m) \\ \text{sn}(x|m) \end{pmatrix}, 
\end{align}
where $A_1,~A_2$ are arbitrary constants (one of them can be fixed from normalization); due to this, the zero-energy state is two-fold degenerate. Fig. \ref{Lame states} shows normalized probability densities of the band-edge states $\psi_{E_0},~\psi_{E_{\sqrt{1-m}}},~\psi_{E_1}$ (black, green, and yellow lines, respectively) in units of $\ell_B$. The three probability densities are periodic, with a period $T=K(m)$. The vertical lines represent multiples of such a period $T$. To plot, we used the modulus $m=1/2$ and the constants $A_1=A_2=1$.   

\section{Quasiperiodic magnetic superlattices supporting bound states} \label{Sec QPMSL}
The model presented in the previous section is periodic and has a positive (and negative) spectrum composed of two allowed plus one forbidden band. We will derive two families of magnetic superlattices supporting bound states and keeping their band-structured spectrum. We use a Darboux transformation applied to a Dirac equation of the form \eqref{Dirac Mag} to insert the bound states. For the sake of completeness, we first introduce the Darboux transformation and then present two quasiperiodic magnetic superlattices supporting bound states. 

\subsection{Darboux transformation of a Dirac equation}

To apply Darboux transformations, we take the dimensionless Dirac equation \eqref{Dirac Mag} and write it in the form \eqref{dirac schrodinger}.  We choose one of the decoupled Hamiltonians $h^{\pm}$, let us say $h^-$. To add a bound state, we first define the operator $h_1^+= h^--\epsilon_1$, where $\epsilon_1$ is a real constant called \emph{factorization energy}. Then, we assume that the \emph{intertwining relation} $h_1^- L_1^+=L_1^+ h_1^+$ is fulfilled, where $L_1^\pm$ are known as \emph{intertwining operators} ($L_1^-$ is the adjoint of $L_1^+$). The proposed expressions for the operators $L^\pm_1$ and $h_1^-$ are  
\begin{equation}
    L_1^\pm=\mp\partial_x+W_1(x,\epsilon_1), \quad h_1^-=-\partial_x^2+V_1^-(x,\epsilon_1),
\end{equation}
where $W_1(x, \epsilon_1)$ is called \emph{superpotential}. From the definition of $h_1^+$ and the intertwining relation, the Schr\"odinger potentials take the form 
\begin{equation}
    V_1^+=V^--\epsilon_1, \quad V_1^-=V_1^+-2W_{1x}(x,\epsilon_1).
\end{equation} 
Each potential is related to a Riccati equation, namely, 
\begin{equation}
   V_1^\pm= W_1(x,\epsilon_1)^2\pm W_{1x}(x,\epsilon_1).
\label{riccati}
\end{equation}
We can map such Riccati equations into Schr\"{o}dinger ones through the ansatz $W_1(x,\epsilon_1)=\partial_x \ln(u_{\epsilon_1})$, leading to: 
\begin{equation}
    (-\partial_x^2+V^+_1)u_{\epsilon_1}=0, \quad (-\partial_x^2+V^-_1)u_{\epsilon_1}^{-1}=0.
\end{equation}
The function $u_{\epsilon_1}$ is often referred to as \emph{seed solution}. Since it fulfills a second-order differential equation, it can be written in terms of two linearly independent solutions as  $u_{\epsilon_1}=u_1(x,\epsilon_1)+\eta\, u_2(x,\epsilon_1)$, where in principle, $\eta$ is an arbitrary constant. Once we fix the seed solution $u_{\epsilon_1}$, we can reconstruct a new Dirac Hamiltonian in the form 
\begin{equation}
H_1=-i \sigma_1 \partial_x + W_1 \sigma_2.     
\end{equation}
The zero-energy eigenspinor of $H_1$ is 
\begin{equation} \label{seed theory}
\psi^{(1)}_{\epsilon_1} \propto \left(\begin{matrix}0 \\ u_{\epsilon_1}^{-1}\end{matrix}\right),
\end{equation}
The remaining solutions are expressed in terms of the intertwining operators and the solutions of the initial Dirac equation as follows: 

\begin{equation}
    \psi^{(1)}=\left(\begin{matrix} u_{\epsilon} \\ \frac{i}{\sqrt{\epsilon-\epsilon_1}} L_1^+ u_{\epsilon}\end{matrix}\right),
\end{equation}
 where $u_{\epsilon}$ fulfills the equation $-\partial_x^2u_{\epsilon}+V^-u_{\epsilon}=\epsilon u_{\epsilon}$ for any $\epsilon \neq \epsilon_1$. 
The corresponding vector potential is given by ${\bf A}^{(1)} =(0,A^{(1)}_y,0)$, where $A^{(1)}_y= B_0 \ell_B(-\ell_B k_y+ W_1(x))$, thus the new magnetic field is $B_1(x,\epsilon_1)=B_0 \partial_x W_1(x,\epsilon_1).$ 

\begin{figure}[t]
     \centering
         \includegraphics[width=0.3\textwidth]{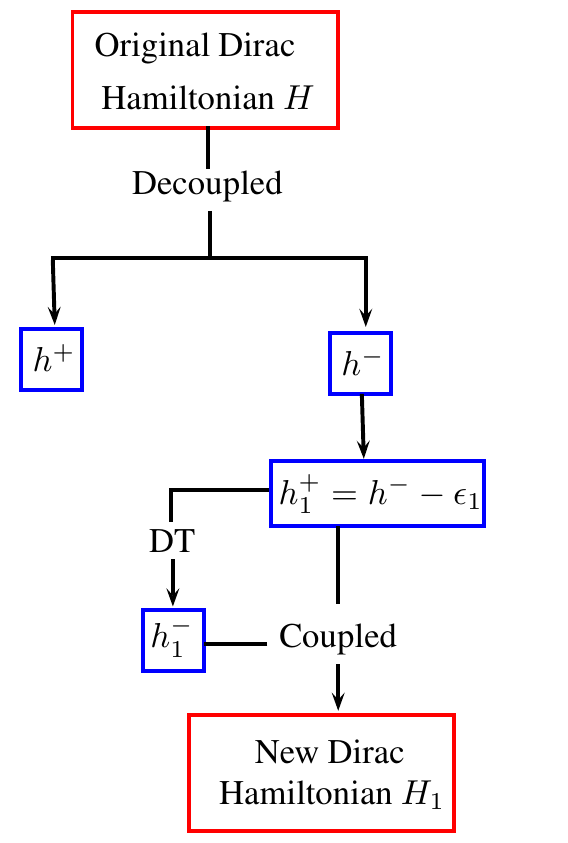}
 \caption{Diagram of the Darboux transformation for a Dirac equation. 
 } \label{map}
\end{figure}

Fig. \ref{map} summarizes the described Darboux transformation. First, we decouple the Dirac Hamiltonian $H$ into two Schrodinger Hamiltonians $H\rightarrow (h^+,h^-)$. Second, we choose the Schr\"{o}dinger Hamiltonian $h^-$ to build two Hamiltonians, $h_1^+=h^--\epsilon_1$ and $h_1^-$, where the last one arises from applying the Darboux transformation to $h^+_1$. Finally, we construct the new Dirac Hamiltonian, $(h_1^+,h_1^-)\rightarrow H_1$. 

\subsection{Quasiperiodic superlattice with an added bound state} \label{QS+1} 

As a first example, we take the Lam\'e magnetic superlattice described in Sec. \ref{sec Lame} and perform a Darboux transformation to add a single bound state. To this end, we select a seed solution $u_{\epsilon_1}=u_1(x,\epsilon_1)+\eta \,u_2(x,\epsilon_1)$, where $u_1$, $u_2$ are the Bloch functions 
\begin{equation}
    u_1=\frac{\sigma (x_0+\omega') \sigma (x+\delta+\omega')}{\sigma (x+\omega') \sigma (x_0+\delta+\omega')} e^{-\zeta (\delta)(x-x_0)}, \quad u_2=\frac{\sigma (x_0+\omega') \sigma (x-\delta+\omega') }{\sigma (x+\omega') \sigma (x_0-\delta+\omega')}e^{\zeta(\delta)(x-x_0)}, \label{1SUSY seed}
\end{equation}
and $\sigma$, $\zeta$ are non-elliptic Weierstrass functions with $\omega'(m)=i K(1-m)$ being the imaginary half-period of $\wp(x)$. Throughout this paper we use $x_0=0$. We take $\eta > 0$ to avoid singularities in the magnetic field. The energy parameter $\epsilon_1$ and the displacement $\delta$ are related by $\epsilon_1=\frac{2}{3}(m+1)-\wp (\delta)-m,$ where $\delta$ is calculated by taking the inverse of $\wp(\delta)$ (see more details in Appendix \ref{Ap Lame}). For this example $\epsilon_1<0$, thus, the spectrum of this system is given by the ground state energy $E^{(1)}_{\epsilon_1}=0$, which emerges as a consequence of the deformation of the external magnetic field, and the allowed energy bands $E^{+(1)}=[\sqrt{-\epsilon_1},\sqrt{1-m-\epsilon_1}]\cup [\sqrt{1-\epsilon_1},\infty)$ and $E^{-(1)}=-E^{+(1)}$. The expression of the bound-state spinor $\psi^{(1)}_{\epsilon_1}$ can be obtained directly by the substitution of \eqref{1SUSY seed} into \eqref{seed theory} while the positive-energy band-edge states become
\begin{equation}
 \psi^{(1)}_{E_0}= \begin{pmatrix} \text{dn}(x|m) \\ \frac{i}{\sqrt{-\epsilon_1}}L_1^+\text{dn}(x|m) \end{pmatrix}, \quad  \psi^{(1)}_{E_{\sqrt{1-m}}}= \begin{pmatrix} \text{cn}(x|m) \\ \frac{i}{\sqrt{1-m-\epsilon_1}}L_1^+\text{cn}(x|m) \end{pmatrix}, \quad \psi^{(1)}_{E_1}=
\begin{pmatrix} \text{sn}(x|m) \\ \frac{i}{\sqrt{1-\epsilon_1}}L_1^+\text{sn}(x|m) \end{pmatrix}.
\end{equation} 
As a final step, the corresponding external magnetic field $B_1(x,\epsilon_1)= B_0 \partial_x W_1 = B_0 \partial_x^2 \ln (u_{\epsilon_1})$ (see appendix~\ref{explicit expression}) is recovered.

Fig. \ref{Energy bands1} (left) shows the probability density $|\psi^{(1)}_{\epsilon_1}|^2$ of the added bound state, which is confined to this region by the aperiodicity in the magnetic field. In Fig. \ref{Energy bands1} (center), we plot the probability density of the band-edge states, where the vertical lines indicate multiples of the initial Lam\'e system's period $\mathcal{T}=2K(m)$. We see that the band-edge states recover their periodicity far from the origin. Fig. \ref{Energy bands1} (right) presents a plot of the generated quasiperiodic magnetic superlattice with an added bound state (red curve) and the initial Lam\'e superlattice (gray shadow) for comparison (see \eqref{lame magnetic}), in units of $B_0$. The magnetic field $B_1$ shows an aperiodicity around the origin introduced by the Darboux transformation, which is the cause for the confined state to appear. An impurity or defect in the superlattice could cause this modification in an experimental setup. As an important remark, the magnetic field $B_1(x,\epsilon_1)$ is actually a biparametric family of quasiperiodic magnetic fields, depending on the parameters $\epsilon_1 < 0$ and $\eta >0$. In the case $\eta=0$ or $\eta\rightarrow\infty$, it returns just a displacement of the Lam\'e magnetic superlattice and the bound state at $E_{\epsilon_1}^{(1)}=0$ disappears.  

\begin{figure}[t]
    \begin{subfigure}[b]{0.32\textwidth}
         \centering
         \includegraphics[width=\textwidth]{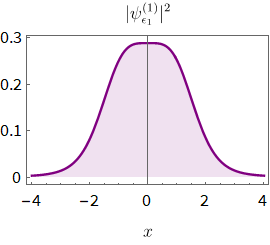}
     \end{subfigure}
     \begin{subfigure}[b]{0.32\textwidth}
         \centering
         \includegraphics[width=\textwidth]{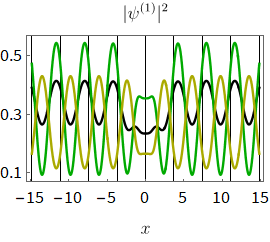}
     \end{subfigure}
     \begin{subfigure}[b]{0.32\textwidth}
         \centering
         \includegraphics[width=\textwidth]{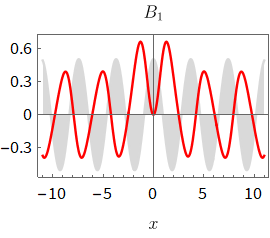}
     \end{subfigure}
 \caption{Probability densities of the added bound state $|\psi^{(1)}_{\epsilon_1}|^2$ (left) and of the band-edge states (center): $|\psi^{(1)}_{E_0}|^2$ (black line), $|\psi^{(1)}_{E_{\sqrt{1-m}}}|^2$ (green line) and $|\psi^{(1)}_{E_1}|^2$ (yellow line). (Right) Quasiperiodic magnetic superlattice with an added bound state (red) and the Lam\'e magnetic superlattice (gray shadow). The used parameters are $m=1/2$, $\eta=1$, $\epsilon_1=-1/2$.} \label{Energy bands1}
\end{figure}

At large values of $x$, the magnetic field $B_1$ recovers its periodicity with an added phase. To see this, note that the functions $u_{1,2}$ grow exponentially in our case of interest $\epsilon_{1} < 0$ and fulfill the relation $u_1(x)u_2(x+\delta)=c$, where $c$ is some constant. We can define an $x_c>0$ such that $u_1(x_c)= c/u_2(x_c+\delta) \sim 0$ and $u_2(-x_c + \delta)= c/u_1(-x_c) \sim 0$. Then, for $x>x_c$, the seed solution $u_{\epsilon_1} \rightarrow u_2$. In the same way, for a negative large value of $x$, ($x<-x_c$), the seed solution $u_{\epsilon_1} \rightarrow u_1$. 
Therefore the superpotential has the following asymptotic behaviors when $\pm x > x_c$:
\begin{equation}
    W_1(x,\epsilon_1) \rightarrow \,m\,\,\text{sn}(x|m)\,\text{sn}(x\pm\delta|m)\,\text{sn}(\mp\delta|m)+\beta_\pm,
\end{equation}
where $\beta_\pm=\zeta (\omega'\pm\delta)-\zeta (\omega')+\zeta (\mp\delta)$ are constants. The superpotential tends to be a periodic function in these regions, $ W_1(x+\mathcal{T},\epsilon_1)= W_1(x,\epsilon_1)$, where $\mathcal{T}=2K(m)$. The above implies the periodicity of the intertwining operator $L^+_1$, the new band-edge states, and the magnetic field outside the origin:
\begin{equation}
     B_1(x,\epsilon_1) \rightarrow B_0(\text{dn}(x\pm\delta|m)^2-\text{dn}(x|m)^2).
\end{equation}

\subsection{Quasiperiodic superlattices with two bound states}\label{2SUSY Lame}

We can iterate the Darboux transformation to add more than one bound state. In particular, the second-order transformation will add two of such states to the superlattice. The second iteration begins once we reconstruct $H_1$. Then we decouple the Dirac equation into two Schr\"odinger equations with Hamiltonians $h_1^\pm$. Now, we take $h^+_2 = h^-_1 - \epsilon_2$ and propose the new intertwining relation $h_2^- L_2^+=L_2^+ h_2^+$.   
The Hamiltonian $h_2^-$ and the intertwining operator $L_2^+$ are given by 
\begin{eqnarray}
L_2^+ = - \partial_x + \frac{\vartheta_x}{\vartheta},    \qquad h_2^-= - \partial_x^2 + V^-_2(x). \label{eq14}
\end{eqnarray}
The seed solution $\vartheta$ of this second transformation satisfies $h_1\vartheta=\epsilon_2\vartheta$, for $\epsilon_2\neq\epsilon_1$, then there is a $u_{\epsilon_2}$ such that $\vartheta=L_1^+ u_{\epsilon_2}$ fulfilling $-\partial_x^2 u_{\epsilon_2} + V^- u_{\epsilon_2} = (\epsilon_1+\epsilon_2) u_{\epsilon_2}$. Thus, the potential $V^-_2$ takes the form
\begin{eqnarray} 
V^-_2= V^+_2 - 2 \partial_x^2(\ln \vartheta)  = V^+_2 -2 \partial_x^2\ln \left(\frac{W(u_{\epsilon_1},u_{\epsilon_2})}{u_{\epsilon_1}}\right), 
\label{eq15}
\end{eqnarray}
where $W(f,g)=f g_x-f_x g$ is the Wronskian of $f$ and $g$. There are now two bound states, one for each first-order transformation. The second-order superpotential and magnetic field are given by 
\begin{equation}
  W_2(x,\epsilon_1,\epsilon_2)= \partial_x \ln\left(\frac{W(u_{\epsilon_1},u_{\epsilon_2})}{u_{\epsilon_1}}\right), \quad   B_2(x,\epsilon_1,\epsilon_2)=B_0 \partial_x \,W_2(x,\epsilon_1,\epsilon_2).
  \label{superpotential2}
\end{equation}
 This second-order transformation produces a quasiperiodic magnetic superlattice supporting two bound states. We start from the Lam\'e magnetic superlattice; the seed solutions that we consider $u_{\epsilon_{j}}=u_1(x,\epsilon_{j})+\eta_{j}u_2(x,\epsilon_{j})$, $j=1,2$, $\epsilon_2<0$, $\epsilon_1<0$ fulfill the Lam\'e equation, see \eqref{1SUSY seed}. 
The system under construction will support two bound states with energies $ E^{\pm(2)}_{\epsilon_1}=\pm \sqrt{-\epsilon_2}, \quad E^{(2)}_{\epsilon_2}=0$. The bound states are given by
\begin{eqnarray} \label{eq18}
\Psi^{(2)}_{\epsilon_1} \propto \left(\begin{matrix} u_{\epsilon_1}^{-1} \\ L_2^+u_{\epsilon_1}^{-1}
\end{matrix}\right), \quad
\Psi^{(2)}_{\epsilon_2} \propto \left(\begin{matrix} 0 \\ \frac{u_{\epsilon_1}}{W(u_{\epsilon_1},u_{\epsilon_2})}
\end{matrix}\right).
\end{eqnarray}
The allowed energy bands in the positive part of the spectrum become 
\begin{equation}
E^{+(2)}=[\sqrt{(-\epsilon_1-\epsilon_2)}, \sqrt{(1-m-\epsilon_1-\epsilon_2)}]\cup [\sqrt{(1-\epsilon_1-\epsilon_2)},\infty),    
\end{equation}
and $E^{-(2)}=-E^{+(2)}$. The positive-energy band-edge states are expressed as 
\begin{align}
 \psi^{(2)}_{E_0}= \begin{pmatrix} \frac{i}{\sqrt{(-\epsilon_1)}}L_1^+\text{dn}(x|m) \\ -\frac{1}{\sqrt{(-\epsilon_1-\epsilon_2)}\sqrt{(-\epsilon_1)}}L_2^+L_1^+\text{dn}(x|m) \end{pmatrix}, \nonumber \\ 
 \psi^{(2)}_{E_{\sqrt{1-m}}}= \begin{pmatrix} \frac{i}{\sqrt{(1-m-\epsilon_1)}}L_1^+\text{cn}(x|m) \\ -\frac{1}{\sqrt{(1-m-\epsilon_1-\epsilon_2)}\sqrt{(1-m-\epsilon_1)}}L_2^+L_1^+\text{cn}(x|m) \end{pmatrix},  \nonumber \\
 \psi^{(2)}_{E_1}= \begin{pmatrix} \frac{i}{\sqrt{(1-\epsilon_1)}}L_1^+\text{sn}(x|m) \\ -\frac{1}{\sqrt{(1-\epsilon_1-\epsilon_2)}\sqrt{(1-\epsilon_1)}}L_2^+L_1^+\text{sn}(x|m) \end{pmatrix}. \label{2susy band a}
\end{align} 

\begin{figure}[t]
    \begin{subfigure}[b]{0.32\textwidth}
         \centering
         \includegraphics[width=\textwidth]{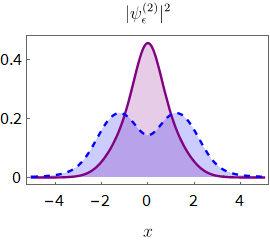}
     \end{subfigure}
     \begin{subfigure}[b]{0.32\textwidth}
         \centering
         \includegraphics[width=\textwidth]{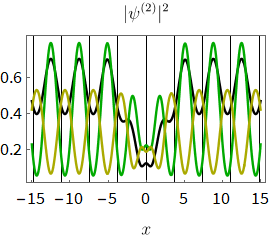}
     \end{subfigure}
    \begin{subfigure}[b]{0.32\textwidth}
         \centering
         \includegraphics[width=\textwidth]{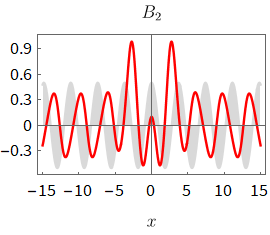}
     \end{subfigure} 
 \caption{Probability density of the two bound states in \eqref{eq18}, $|\Psi_{\epsilon_2}|^2$ (purple line) and $|\Psi_{\epsilon_1}|^2$ (blue dashed line) (left) and of the band-edge states $|\psi^{(2)}_{m}|^2$ (black line), $|\psi^{(2)}_{1}|^2$ (green line) and $|\psi^{(2)}_{1+m}|^2$ (yellow line), with period $\mathcal{T}=2K(m)$ (center). (Right) Quasiperiodic magnetic superlattice with two added bound states (red line) and Lam\'e magnetic superlattice (gray shadow). We use the parameters $\eta_1=1$, $\eta_2=-1$, $\epsilon_1=-1/2$, $\epsilon_2=-1$.} \label{Energy bands2}
\end{figure}

Fig. \ref{Energy bands2} shows the probability densities of the described states: on the left for the bound states \eqref{eq18}, on the center for the band-edge states \eqref{2susy band a}. 
The corresponding magnetic field, calculated using equation \eqref{superpotential2}, is plotted on the right (red curved) besides the Lam\'e magnetic field (gray shadow) for comparison, in units of $B_0$. Similar to the quasiperiodic magnetic superlattice with a single bound state, the period of the magnetic field far from the origin is $\mathcal{T}=2K(m)$. The deformations in the magnetic field now increase in magnitude with respect to the first transformation. 

\section{Domain walls}

When a graphene layer is deposited on a substrate, like boron nitrate, the behavior of the charge carriers can be modeled by the Dirac equation with an effective mass,
\begin{equation}
  {\cal H} \Phi = E \Phi, \quad 
     {\cal H} = \hbar v_F(-i\sigma_1\partial_x+\sigma_2 k_y) +\mu(x) v_F^2\sigma_3 .
     \label{originalH with mass}
\end{equation}
where the mass term represents a break of the symmetry between the two triangular sublattices of graphene. A domain wall occurs when there is a change in the symmetry breaking of the triangular sublattices, which is modeled by a position-dependent mass term $\mu=\mu(x)$ fulfilling \cite{Semenoff08}  
\begin{equation} \label{con1}
   \lim_{x\rightarrow -\infty} \mu(x)=-\mu < 0, \quad \lim_{x\rightarrow \infty} \mu(x)=\mu> 0. 
\end{equation}

We can use the results of Sec. \ref{sec Lame} and \ref{Sec QPMSL} to obtain different mass profiles $\mu(x)$ having domain walls and the corresponding solutions of the Dirac equation.  Since the proposed mass profile depends only on one coordinate, we can use separation of variables to simplify the problem as follows, $\Phi(x,y)=e^{-ik_yy}X(x)$. Then we introduce the unitary transformation ${\cal R}=\exp(i\frac{\pi}{4}\sigma_1)$, such that $H={\cal R} {\cal H}{\cal R}^{-1}$ and $\psi(x)={\cal R} X(x)$. Therefore, $H$ acquires a form similar to equation \eqref{Dirac Mag}, and the Darboux transformation can be as well applied. To obtain a dimensionless Hamiltonian, the variables are renamed as $x\rightarrow x/l_{\mu}$, $k_y\rightarrow l_{\mu}k_y$ and $E\rightarrow l_{\mu}E/\hbar v_F$, where  $l_{\mu}= \hbar / v_F\mu_0$ is the  characteristic length, similar to the magnetic case for which $\mu_0$ represents the mass amplitude $\mu(x)=\mu_0 W_0(x)$ (see section~\ref{sec Lame}). This leads to the Dirac equation
\begin{equation}
    (- i \sigma_1\partial_x+W_0(x)\sigma_2+k_y\sigma_3)\psi=E\psi.
    \label{dirac H equation02}
\end{equation}
This is analog to equation \eqref{Dirac Mag}, where the momentum $k_y$ in the rotated system can be interpreted as an extra mass term. We can also write \eqref{dirac H equation02} in terms of its components $\psi=(\psi^+,\psi^-)^T$, in order to obtain instead of \eqref{Dirac Mag} and \eqref{dirac schrodinger},  

\begin{equation}
    \psi^\pm=-\frac{i}{E\pm k_y}\left(\pm \partial_x+W_0(x)\right)\psi^\pm, \quad  -\psi^{\mp}_{xx}+\left(W_0(x)^2\pm W_{0x}(x)\right)\psi^\mp=\left(E^2-k_y^2\right)\psi^\mp.
\end{equation}
We now use the Lam\'e magnetic field \eqref{lame magnetic} and take the limit when the period of oscillation goes to infinity, $m\rightarrow 1$. Then: 
\begin{equation}
\lim_{m\rightarrow 1} W_0(x)= \tanh(x).
\end{equation}
There is a single bound state $\psi_{E_0}=(0,\text{sech}(x))^T$, with associate energy $ E=|k_y|$. 

\begin{figure}[t]
 \begin{subfigure}[b]{0.32\textwidth}
         \centering
         \includegraphics[width=\textwidth]{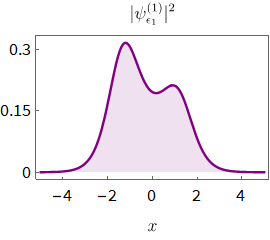}
    \end{subfigure}
     \begin{subfigure}[b]{0.32\textwidth}
         \centering
         \includegraphics[width=\textwidth]{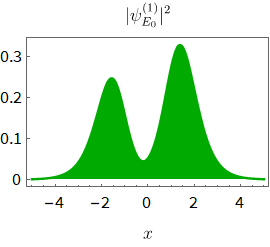}
    \end{subfigure}
    \begin{subfigure}[b]{0.32\textwidth}
         \centering
         \includegraphics[width=\textwidth]{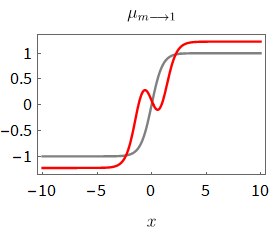}
     \end{subfigure}
 \caption{(Left) Bound state generated from the modification of the mass profile at zero energy $|\psi^{(1)}_{\epsilon_1}|^2$. (Center) Bound state $|\psi^{(1)}_1|^2$, which originates from the band collapse when $m\rightarrow 1$. (Right) Modified mass profile from the first-order Darboux transformation in the limit $m\rightarrow 1$ (red line) as compared with the initial mass profile (gray line). We have taken $\eta=2/3$, $\epsilon_1=-1/2$, $k_y=1$.} \label{mass limit01}
\end{figure}

Moreover, we can use the results of Sec. \ref{Sec QPMSL} to generate new domain walls. For example, we can take the quasiperiodic magnetic superlattice with an added bound state, take the limit $m \rightarrow 1$ and apply at the end the unitary rotation $\cal R$. The mass amplitude is then $\mu(x) = \mu_0 W_1(x,\epsilon)$, where 
\begin{equation}
   W_1(x,\epsilon_1)=\tanh (x)+\left(\frac{\text{csch}(\delta) \cosh (x) \left(\eta\,e^{2 x \coth (\delta)}-1\right)}{\cosh (\delta+x)+\eta \cosh (x-\delta) e^{2 x \coth (\delta)}}-\frac{4\tanh (x/2)}{1+\tanh^2(x/2)}\right),
\end{equation}
with $\epsilon_1=1/3-\wp (\delta)$, $\eta> 0$. This mass profile fulfills the domain wall conditions, as it is shown in Fig. \ref{mass limit01}. Also, it is presented the probability density associated with the generated bound states $\psi^{(1)}_{E_0}= \psi^{(1)}_{E_{\sqrt{1-m}}}$ and $\psi^{(1)}_{\epsilon_1}$. The first bound state arises from the collapse of the band-edge states, with energy $E^{(1)}=\sqrt{k_y^2-\epsilon_1}$, the second appears from the mass term modification, which is a consequence of the Darboux transformation with energy $E^{(1)}_{\epsilon_1}=0$. The corresponding eigenspinors have the explicit form   
\begin{align}
 \psi^{(1)}_{E_0,E_{\sqrt{1-m}}}= \begin{pmatrix} \text{sech}(x) \\ \frac{i}{\sqrt{k_y^2-\epsilon_1}-k_y} L_1^+\text{sech}(x) \end{pmatrix}, \\
 \psi^{(1)}_{\epsilon_1}= \begin{pmatrix} 0 \\  \frac{\cosh (\delta) \cosh (x) e^{x \coth (\delta)}}{\cosh (\delta+x)+\eta_1 \cosh (x-\delta) e^{2 x \coth (\delta)}} \end{pmatrix},
\end{align} 
where the intertwining operator is given by $L_1^+=-\partial_x+W_1(x,\epsilon_1)$. Different mass profiles with domain walls can be constructed taking the limit $T\rightarrow \infty$ of the quasiperiodic magnetic superlattices generated through the technique presented in this section.  

\section{Summary}

In this work, we have presented a model that describes a monolayer graphene interacting with a Lam\'e magnetic superlattice which is periodic and smooth in the whole domain. The system has a band-structured spectrum with its eigenspinors having a closed explicit form.  

Moreover, we have shown how to include magnetic field deformations by adding discrete energy levels to the spectrum without destroying the solvability of the system. The resulting magnetic fields turn out to be quasiperiodic, in the sense that far from the localized deformations, they become asymptotically periodic. The technique allows iterations; thus, adding any finite number of discrete energies to the spectrum turns out to be possible. We exemplified the technique through the simplest Lam\'e magnetic superlattice.  

It was shown that the topics of magnetic superlattices and domain walls in gapped graphene can be connected if we first take the limit when the period of the superlattice goes to infinity and then a unitary rotation to the Dirac equation is applied. This connection also holds for the quasiperiodic magnetic superlattices constructed through Darboux transformations. The energy levels in the discrete spectrum keep their nature in the domain wall limit, but some extra levels could arise due to the collapse of energy bands. The deformations of the magnetic field translate into deformations of the mass term in gapped graphene.

\section*{Acknowledgments}
The authors acknowledge Consejo Nacional de Humanidades Ciencia y Tecnolog\'{\i}a (CONAHCyT-M\'exico) under grant FORDECYT-PRONACES/61533/2020.

\appendix
\section{Schr\"odinger equation with Lam\'e potentials} \label{Ap Lame}
For the sake of completeness, we summarize the solution of Schr\"odinger equation with Lam\'e potentials in this appendix; the reader can find definitions of the involved functions in \cite{Abramowitz64,Erdelyi2}, a review of the problem in \cite{Whittaker50}, and a complete pedagogical approach in \cite{Arscott64}. First, we start from a Schr\"odinger Hamiltonian $h=-\partial_x^2 + V(x)$ where the potential $V(x)$ is periodic, nonsingular, with period $\mathcal{T}$. Thus, the solutions of the Schrodinger equation $h \psi = E \psi$ fulfill the Bloch theorem,
\begin{equation}
    \psi(x)=e^{i\kappa x} u_\kappa (x),
\end{equation}
where $u_\kappa$ is a periodic function $u_\kappa (x+\mathcal{T})=u_\kappa (x)$ and $\kappa$ denotes the momentum of the crystal. Note also that $\psi(x+\mathcal{T})=e^{i\kappa \mathcal{T}}\psi(x)$. The spectrum consists of allowed energy bands with edges such that $\kappa\,\mathcal{T} =0,\pi$, then the band edge eigenfunctions fulfill $\psi(x + \mathcal{T}) = \pm \psi(x)$. 

Now, the Lam\'e potentials are given by 
\begin{equation}
    V(x)=n(n+1)\,m\, \text{sn}(x|m)^2,
\end{equation}
where $m$ is the modulus of the Jacobi elliptic function $\text{sn}(x|m)$ such that $0 \leq m \leq 1$, and $n$ is a natural number. The period of the potential is $\mathcal{T}=2K(m)$, where 
\begin{equation}
    K(m)=\int_0^{\pi/2}\frac{d\theta}{\sqrt{1-m \sin^2\theta}}.
\end{equation} 
This system has $2n+2$ bands, $n+1$ allowed plus $n+1$ forbidden bands. An equivalent expression in terms of the Weierstrass elliptic function $\wp(x)$ is given by
\begin{equation}
    V(x)=n(n+1)\left(\wp(x+iK(1-m))+\frac{1}{3}(m+1)\right).
\end{equation}
We are using the short notation $\wp(x,g_2,g_3)\equiv \wp(x)$, and in general the Weierstrass functions $\sigma(x,g_2,g_3)\equiv \sigma(x)$ and $\zeta(x,g_2,g_3)\equiv \zeta(x)$ in the variable $x$ for parameters, or elliptic invariants, $g_2$, and $g_3$. In this case 
\begin{equation}
    g_2(m)=\frac{4}{3} \left(m^2-m+1\right), \quad g_3(m)=\frac{4}{27} (m-2) (2 m-1) (m+1).
\end{equation}

In this paper we will consider $n=1$, the simplest case of the Lam\'e potentials. The solution of the stationary Schr\"{o}dinger equation $-\psi_{xx}+V(x)\psi=\epsilon\psi$ for any value of the energy parameter $\epsilon$ is given by 
\begin{align}
    \psi=u_1(x,\epsilon)+\eta \,u_2(x,\epsilon),
\end{align}
where $\eta$ is an arbitrary constant and $u_1$, $u_2$ are the Bloch functions 
associated to $\epsilon$ given by 
\begin{equation}
    u_1(x,\epsilon)=\frac{\sigma (x_0+\omega') \sigma (x+\delta+\omega')}{\sigma (x+\omega') \sigma (x_0+\delta+\omega')} e^{-\zeta (\delta)(x-x_0)}, \quad u_2(x,\epsilon)=\frac{\sigma (x_0+\omega') \sigma (x-\delta+\omega') }{\sigma (x+\omega') \sigma (x_0-\delta+\omega')}e^{\zeta(\delta)(x-x_0)},
\end{equation}
with $\omega'(m)=i K(1-m)$ being the imaginary half-period of $\wp(x)$ and $x_0$ a reference point in the $x$ axis.
 The energy parameter $\epsilon$ and the displacement $\delta$ are related by $\epsilon=2(m+1)/3-\wp (\delta)$, \textit{i.e.},
the parameter $\delta$ is calculated through the inverse of $\wp(\delta)$.
The functions $u_1$, $u_2$ fulfill the traditional Bloch relation 

\begin{equation}
    u_1(x+\mathcal{T})=e^{i\kappa \mathcal{T}} u_1(x), \quad u_2(x+\mathcal{T})=e^{-i\kappa\mathcal{T}} u_2(x), 
\end{equation}
where $\kappa=2i(\omega\zeta(\delta)-\delta\zeta(\omega))$. The band-edge states are given in terms of the Jacobi elliptic functions as follows 
\begin{equation}
    \psi_m = \text{dn}(x|m),\quad  \psi_1 = \text{cn}(x|m), \quad  \psi_{1+m}=\text{sn}(x|m).
\end{equation}
Their corresponding energies are $E=m,~1,~1+m.$ Therefore, the spectrum is composed by the allowed bands $[m,1]$, $[1+m,\infty)$, and the gaps $(-\infty,m)$, $(1,1+m)$.

\section{Magnetic field expression for the quasiperiodic superlattice with an added bound state
}\label{explicit expression}

The explicit expressions of the designed magnetic fields are, in general, lengthy. Here we present the corresponding expression for the subsection \ref{QS+1}: 
\begin{equation}
    B_1(x,\epsilon_1)=B_0\partial_x W_1(x,\epsilon_1),
\end{equation}
 where $W_1=\frac{\partial_x(u_1+\eta\,u_2)}{u_1+\eta\, u_2}$ is the superpotential, 
and $u_1$, $u_2$ are the Bloch functions (see appendix~\ref{Ap Lame}). Then, the derivative of the superpotential is  
\begin{equation*}
    \partial_x  W_1(x,\epsilon_1)=\frac{\partial_{xx}u_1+\eta\,\partial_{xx}u_2}{u_1+\eta\, u_2}-\left(\frac{\partial_xu_1+\eta\,\partial_xu_2}{u_1+\eta\, u_2}\right)^2.
\end{equation*}
The explicit form of the derivatives are 
\begin{align*}
&\partial_xu_1= -\frac{e^{-x \zeta(\delta)}\sigma(\omega') \sigma(x+\delta+\omega')}{\sigma(\delta +\omega') \sigma(x+\omega')} \left(\zeta(\delta)-\zeta(x+\delta+\omega')+\zeta(x+\omega')\right),\\
& \partial_xu_2=-\frac{e^{x \zeta(\delta)}\sigma(\omega') \sigma(x-\delta+\omega')}{\sigma(\delta-\omega') \sigma(x+\omega')} \left(\zeta(\delta)+\zeta(x-\delta+\omega')-\zeta(x+\omega')\right),
\end{align*}
\begin{align*}
 & \partial_{xx}u_1=\frac{e^{-x \zeta (\delta)}\sigma(\omega') \sigma (x+\delta +\omega') \left(\wp (x+\omega')-\wp (x+\delta +\omega')+(\zeta (\delta)-\zeta (x+\delta +\omega')+\zeta (x+\omega'))^2\right)}{\sigma (\delta +\omega') \sigma (x+\omega')},\\   
 & \partial_{xx}u_2=-\frac{e^{x \zeta (\delta)} \sigma (\omega') \sigma (x-\delta +\omega') \left(\wp (x+\omega')-\wp (x-\delta +\omega')+(\zeta (\delta)+\zeta (x-\delta +\omega')-\zeta (x+\omega'))^2\right)}{\sigma (x+\omega') \sigma (\delta -\omega')}.
\end{align*}
Therefore, the magnetic field is given by 

\begin{align}
 B_1(x,\epsilon_1)=B_0 \left(\wp \left(x+\omega '\right)-\frac{P_1+P_2+P_3}{\left(\eta  e^{2 x \zeta(\delta)} \sigma\left(\delta+\omega '\right) \sigma\left(x-\delta+\omega '\right)-\sigma\left(\delta-\omega '\right) \sigma\left(x+\delta+\omega '\right)\right)^2}\right),
\end{align}
where 
\begin{align*}
     P_1 &=\eta ^2 \wp \left(x-\delta+\omega '\right) e^{4 x \zeta(\delta)} \sigma\left(\delta+\omega '\right)^2 \sigma\left(x-\delta+\omega '\right)^2,\\
     P_2 &= \eta  e^{2 x \zeta(\delta)} \sigma\left(\delta-\omega '\right) \sigma\left(\delta+\omega '\right) \sigma\left(x-\delta+\omega '\right) \sigma\left(x+\delta+\omega '\right) \\
    &\left(\left(2 \zeta(\delta)+\zeta\left(x-\delta+\omega '\right)-\zeta\left(x+\delta+\omega '\right)\right)^2-\wp \left(x-\delta+\omega '\right)-\wp \left(x+\delta+\omega '\right)\right),\\
    P_3 &=\wp \left(x+\delta+\omega '\right) \sigma\left(\delta-\omega '\right)^2 \sigma\left(x+\delta+\omega '\right)^2.
\end{align*}



\begin{thebibliography}{99}
\bibitem{Wallace1947} P.R. Wallace. The band theory of graphite, {\it Phys. Rev.} {\bf 71}, 622 (1947).
\bibitem{CastroNeto2009} A. H. Castro Neto, F. Guinea, N. M. R. Peres, K. S. Novoselov, and A. K. Geim. The electronic properties of graphene, {\it Rev. Mod. Phys.} {\bf 81}, 109 (2009).
\bibitem{Semenoff1984} G.W. Semenoff. Condensed-Matter Simulation of a Three-Dimensional Anomaly, {\it Phys. Rev Lett.} {\bf 53}, 2449 (1984).
\bibitem{Novoselov2005} Novoselov, K S and Geim, A K and Morozov, S V and Jiang, D and Katsnelson, M I and Grigorieva, I V and Dubonos, S V and Firsov, A A. Two-dimensional gas of massless {Dirac} fermions in graphene, {\it Nat.} {\bf 438}, 197 (2005).
\bibitem{Katsnelson2006} M. I. Katsnelson, K. S.  Novoselov, and A. K. Geim. Chiral tunnelling and the Klein paradox in graphene, {\it Nat. Phys.} {\bf 2}, 620 (2006).
\bibitem{stander2009} N. Stander, B. Huard, and D. Goldhaber-Gordon. Evidence for Klein Tunneling in Graphene p-n Junctions, {\it Phys. Rev. Lett.} {\bf 102}, 026807 (2009). 
\bibitem{Young2009} A.F. Young and P. Kim. Quantum interference and Klein tunnelling in graphene heterojunction, {\it Nat. Phys.} {\bf 5}, 222 (2009).
\bibitem{Katsnelson2006b} M.I. Katsnelson. Zitterbewegung, chirality, and minimal conductivity in graphene, {\it Eur. Phys. J. B}. {\bf 51}, 157 (2006).
\bibitem{Rusin2008} T.M. Rusin and W. Zawadzki. Zitterbewegung of electrons in graphene in a magnetic field, {\it Phys. Rev. B} {\bf 78}, 125419 (2008).
\bibitem{Allor2008} D. Allor, T.D. Cohen, and D.A. McGady. Schwinger mechanism and graphene, {\it Phys. Rev. D} {\bf 78}, 1 (2008). 
\bibitem{Fillion-Gourdeau2015} F. Fillion-Gourdeau and S. MacLean. Time-dependent pair creation and the Schwinger mechanism in graphene, {\it Phys. Rev. B} {\bf 92}, 1 (2015).
\bibitem{Kronig1931} R. de L. Kronig and W.G. Penney. Quantum mechanics of electrons in crystal lattices, {\it Proc. Royal Society of London} {\bf 130}, 499 (1931).
\bibitem{Park2008} C.-H. Park, L. Yang, Y-W Son, M.L. Cohen, and S.G. Louie. Anisotropic behaviours of massless {Dirac} fermions in graphene under periodic potentials, {\it Nat. Phys.} {\bf 4}, 213 (2008).
\bibitem{Barbier2008} M. Barbier, F.M. Peeters, P. Vasilopoulos, and J.M. Pereira. Dirac and Klein-Gordon particles in one-dimensional periodic potentials, {\it Phys. Rev. B} {\bf 77}, 115446 (2008).
\bibitem{Park2009} C.H. Park, Y.W. Son, L. Yang, M.L. Cohen, and S.G. Louie. Landau levels and quantum hall effect in graphene superlattices, {\it Phys. Rev. Lett.} {\bf 103}, 1 (2009). 
\bibitem{Brey2009} L. Brey and H.A. Fertig. Emerging zero modes for graphene in a periodic potential, {\it Phys. Rev. Lett.} {\bf 103}, 1 (2009).
\bibitem{ibrapeet95} I.S. Ibrahim and F.M. Peeters. Two-dimensional electrons in lateral magnetic superlattices, {\it Phys. Rev. B}. {\bf 52}, 17321 (1995).
\bibitem{life06} J-F Liu, W-J Deng, K. Xia, C. Zhang, and Z. Ma. Transport of spin-polarized electrons in a magnetic superlattice, {\it Phys. Rev. B} {\bf 73}, 155309 (2006).
\bibitem{wu2008} Q-S Wu, S-N Zhang, and S-J Yang. Transport of the graphene electrons through a magnetic superlattice, {\it J. Phys. Condens. Matt.} {\bf 20}, 485210 (2008).
\bibitem{DellAnna11} L. Dell\'{\,}Anna and A. De Martino. Magnetic superlattice and finite-energy Dirac points in graphene, {\it Phys. Rev. B} {\bf 83}, 1 (2011).  
\bibitem{Maksimova2012} G. M. Maksimova, E.S. Azarova, A.V. Telezhnikov, and V.A. Burdov. Graphene superlattice with periodically modulated Dirac gap, {\it Phys. Rev. B} {\bf 86}, 205422 (2012). 
\bibitem{Correa13} F. Correa and V. Jakubsk{\'{y}}. Finite-gap twists of carbon nanotubes and an emergent hidden supersymmetry. {\it Phys. Rev. B} {\bf 87}, 085019 (2013).
\bibitem{Semenoff08} G.W. Semenoff, V. Semenoff, and F. Zhou. Domain Walls in Gapped Graphene, {\it Phys. Rev Lett.} {\bf 101}, 087204 (2008). 
\bibitem{gafi21} G.C. Paul, SK. F. Islam, P. Dutta, and A. Saha. Signatures of interfacial topological chiral modes via RKKY exchange interaction in Dirac and Weyl systems, {\it Phys. Rev. B} {\bf 103}, 115306 (2021).  

\bibitem{nataf20} G.F. Nataf, M. Guennou, J.M. Gregg, D. Meier, J. Hlinka, E.K.H Salje, and J. Kreisel. Domain-wall engineering and topological defects in ferroelectric and ferroelastic materials, {\it Nat. Rev. Phys.} {\bf 2}, 634 (2020).
\bibitem{ghosh2021} R. Ghosh. Position-dependent mass Dirac equation and local Fermi velocity, {\it J. Phys. A} {\bf 55}, 015307 (2021).
\bibitem{roy2021} A. Schulze-Halberg and P. Roy. Dirac systems with magnetic field and position-dependent mass: Darboux transformations and equivalence with generalized Dirac oscillators, {\it Ann. Phys.} {\bf 431}, 168534 (2021).

\bibitem{schulze2022} A. Schulze-Halberg. Darboux transformations for Dirac equations in polar coordinates with vector potential and position-dependent mass, {\it Eur. Phys. J. Plus} {\bf 137}, 832 (2022).

\bibitem{Dunne1998} G. Dunne and J. Feinberg. Self-isospectral periodic potentials and supersymmetric quantum mechanics, {\it Phys. Rev. D} {\bf 57}, 1271 (1998). 

\bibitem{Khare1999} A. Khare and U. Sukhatme. New solvable and quasiexactly solvable periodic potentials, {\it J. Math. Phys.} {\bf 40}, 5473 (1999).  

\bibitem{fnn00} D.J. Fern{\'a}ndez C., J. Negro, and L.M. Nieto. Second-order supersymmetric periodic potentials, {\it Phys. Lett. A} {\bf 275}, 338 (2000).

\bibitem{Fernandez2002b} D.J. Fern{\'{a}}ndez C., B. Mielnik, O. Rosas-Ortiz, and B.F. Samsonov. Nonlocal supersymmetric deformations of periodic potentials, {\it J. Phys. A} {\bf 35}, 309 (2002).

\bibitem{Fernandez2002} D.J. Fern\'andez C., B. Mielnik, O. Rosas-Ortiz, and B.F. Samsonov. The phenomenon of Darboux displacements, {\it Phys. Lett. A} {\bf 294}, 168 (2002). 

\bibitem{fg05} D.J. Fern{\'a}ndez C. and A. Ganguly. New supersymmetric partners for the associated Lam{\'e} potentials, {\it Phys. Lett. A} {\bf 238}, 203 (2005).

\bibitem{fg07} D.J. Fern{\'a}ndez C. and A. Ganguly. Exactly solvable associated Lam{\'e} potentials and supersymmetric transformations, {\it Ann. Phys.} {\bf 322}, 1143 (2007).

\bibitem{bermudez2012} D. Bermudez, D.J. Fern{\'a}ndez C., and N. Fern{\'a}ndez-Garc{\'i}a. Wronskian differential formula for confluent supersymmetric quantum mechanics, {\it Phys. Lett. A} {\bf 376}, 692 (2012).

\bibitem{bermudez2013} A. Berm{\'u}dez, O. Dom{\'\i}nguez, D. G{\'o}mez, and P. Salgado. Finite element approximation of nonlinear transient magnetic problems involving periodic potential drop excitations, {\it Comp. \& Math. Appl.} {\bf 65}, 1200 (2013). 

\bibitem{kuru2009} \c{S}. Kuru, J. Negro, L.M. Nieto. Exact analytic solutions for a Dirac electron moving in graphene under magnetic fields. {\it J. Phys.: Condens. Matter.} {\bf 21}, 455305 (2009).

\bibitem{yesilta2012} {\"O}. Ye{\c{s}}ilta{\c{s}}. Symmetric Hamiltonian model and Dirac equation in 1+ 1 dimensions, {\it J. Phys. A} {\bf 46}, 015302 (2012). 

\bibitem{midya2014} B. Midya and D.J. Fern{\'a}ndez C. Dirac electron in graphene under supersymmetry generated magnetic fields, {\it J. Phys. A} {\bf 47}, 285302 (2014).

\bibitem{Contreras-Astorga2014} A. Contreras-Astorga and A. Schulze-Halberg. The confluent supersymmetry algorithm for Dirac equations with pseudoscalar potentials. {\it J. Math. Phys.} {\bf 55}, 103506 (2014).

\bibitem{junker2020}G. Junker. Supersymmetric Dirac Hamiltonians in (1+1) dimensions revisited. {\it Eur. Phys. J. Plus} {\it 135} 464 (2020).

\bibitem{celeita2020}  M. Castillo-Celeita and D.J. Fernandez C. Dirac electron in graphene with magnetic fields arising from first-order intertwining operators. {\it J. Phys. A} {\bf 53}, 035302 (2020).

\bibitem{bagchi2021} B. Bagchi and R. Ghosh. Dirac Hamiltonian in a supersymmetric framework, {\it J. Phys. A} {\bf 62}, 072101 (2021).

\bibitem{kizilirmak2021} D.D. K{\i}z{\i}l{\i}rmak, {\c{S}}. Kuru, and J. Negro, Dirac-like Hamiltonians associated to Schr{\"o}dinger factorizations, {\it Eur. Phys. J. Plus} {\bf 136}, 1 (2021).

\bibitem{schulze2021} A. Schulze-Halberg. First-order Darboux transformations for Dirac equations with arbitrary diagonal potential matrix in two dimensions, {\it Eur. Phys. J. Plus} {\bf 136}, 1 (2021).

\bibitem{raya2010} A. Raya and E. Reyes. Fermion condensate and vacuum current density induced by homogeneous and inhomogeneous magnetic fields in (2+ 1) dimensions, {\it Phys. Rev. D} {\bf 82}, 016004 (2010).

\bibitem{naba17}  G. Naumis, S. Barraza-Lopez, M. Oliva-Leyva, and H. Terrones. Electronic and optical properties of strained graphene and other strained 2D materials: a review. {\it Reports on Progress Phys.} {\bf 80}, 096501 (2017).

\bibitem{diaz2020} E. D{\'\i}az-Bautista. Schr{\"o}dinger-type 2\textsc{D} coherent states of magnetized uniaxially strained graphene, {\it J. Math. Phys.} {\bf 61}, 102101 (2020).

\bibitem{yesilta2022} {\"O}. Ye{\c{s}}ilta{\c{s}} and J. Furtado. Pseudo-Hermitian Dirac operator on the torus for massless fermions under the action of external fields, {\it Int. J. Mod. Phys. A} {\bf 37}, 2250073 (2022).

\bibitem{Abramowitz64} M. Abramowitz and I.A. Stengun.{\it Handbook of mathematical functions: with formulas, graphs, and mathematical tables} (Dover publications, New York City, 1964).

\bibitem{Erdelyi2} A. Erdelyi. {\it Higher Transcendental Functions} (McGraw-Hill Book Company, Inc. Vol. {\bf 2}, New York, 1953)

\bibitem{Whittaker50} E.T. Whittaker, and G.N. Watson. {\it A Course of Modern Analysis}, Cambridge Mathematical Library (Cambridge University Press, 1950).

\bibitem{Arscott64} F.M. Arscott. {\it Periodic differential equations : an introduction to {M}athieu, {Lam\'e} and {A}llied functions} (Pergamon Press, Oxford, 1964).


\end{thebibliography}
\end{document}